\documentclass[useAMS,usenatbib]{mn2e}
\usepackage{epsfig}
\usepackage{amsmath, amssymb,bm}

\title[Evolution of the Snow Line ]{On the Evolution of the Snow Line in Protoplanetary Discs}

\author[R. G. Martin \& M. Livio]{Rebecca G. Martin and Mario
  Livio \\ Space Telescope Science Institute, 3700 San Martin
  Drive, Baltimore, MD 21218, USA \\ }

\begin{document}

\date{}

\pagerange{\pageref{firstpage}--\pageref{lastpage}} 
\pubyear{2012}
\maketitle

\label{firstpage}

\begin{abstract}
We model the evolution of the snow line in a protoplanetary disc. If
the magneto-rotational instability (MRI) drives turbulence throughout
the disc, there is a unique snow line outside of which the disc is
icy.  The snow line moves closer to the star as the infall accretion
rate drops. Because the snow line moves inside the radius of the
Earth's orbit, the formation of our water-devoid planet is difficult
with this model. However, protoplanetary discs are not likely to be
sufficiently ionised to be fully turbulent. A dead zone at the
mid-plane slows the flow of material through the disc and a steady
state cannot be achieved. We therefore model the evolution of the snow
line also in a time-dependent disc with a dead zone. As the mass is
accumulating, the outer parts of the dead zone become self
gravitating, heat the massive disc and thus the outer snow line does
not come inside the radius of the Earth's orbit, contrary to the fully
turbulent disc model. There is a second, inner icy region, within the
dead zone, that moves inwards of the Earth's orbit after a time of
about $10^6\,\rm yr$. With this model there is sufficient time and
mass in the disc for the Earth to form from water-devoid planetesimals
at a radius of $1\,\rm AU$. Furthermore, the additional inner icy
region predicted by this model may allow for the formation of giant
planets close to their host star without the need for much migration.
\end{abstract}


\section{Introduction}

Bodies in our solar system show a distribution of water abundance. The
innermost terrestrial planets contain little water compared with the
outer planets. Planet formation is thought to occur from planetesimals
that formed in the solar nebula. Within the solar nebula, ice forms
beyond a radius from the central star known as the snow line, $R_{\rm
  snow}$. This is thought to play an important role in the composition
of forming planets. The solid mass density outside the snow line is
much higher because of water ice condensation. In fact, for a solar
composition, water ice abundance is as high as silicate and iron
\citep[e.g.][]{pollack94}.  Observations of the asteroid belt, located between
Mars and Jupiter, suggest that the snow line is currently located
within this region.  The outer asteroids are icy C-class objects
\citep[e.g.][]{abe00,morbidelli00} whereas the inner asteroid belt is
largely devoid of water. This implies that when planetesimal formation
occurred the snow line was located at around $R_{\rm snow}=2.7\,\rm
AU$ from the Sun.

The snow line occurs at a temperature, $T_{\rm snow}$, that is in the
range of $145\,\rm K$ \citep{podolak04} to $170\,\rm K$
\citep{hayashi81}, depending on the partial pressure of nebular water
vapour.  Its distance from the star is usually calculated in a fully
turbulent steady state accretion disc
\cite[e.g.][]{sasselov00,lecar06,kennedy06,kennedy08}. The
magneto-rotational instability (MRI) is thought to drive the
turbulence in the disc \citep{balbus91} and thus transport angular
momentum outwards allowing accretion on to the central star
\citep[e.g.][]{fromang06}. Recent calculations have used a detailed
disc structure and include a stellar radiation flux and viscous
dissipation of the gas as the main heating sources. The disc is
assumed to be in a quasi-steady state as the infall accretion rate
decreases in time \citep[e.g.][]{ida05,garaud07,min11}.
\cite{davis05} found that the radius of the snow line reaches a
minimum of about $0.6\,\rm AU$, inside the current orbit of the Earth
and this was confirmed by \cite{garaud07}. \cite{garaud07} find that
the snow line migrates inwards as the accretion rate drops (down to an
accretion rate of around $\dot M=10^{-10}\,\rm M_\odot\,yr^{-1}$)
because the viscous dissipation decreases. However, as the accretion
rate drops below $10^{-10}\,\rm M_\odot\,yr^{-1}$, the snow line
migrates outwards again as the disc becomes optically thin and the
temperature rises.

The fully turbulent steady state model predicts a minimum snow line
radius within the Earth's orbit. If planetesimals formed during this
time, the Earth would have formed from icy bodies. This appears to be
contradictory with the current water content on Earth that is very low
at around $0.023\%$ by weight \citep{lewis04}. For comparison, the
outer solar system planets have a mass fraction of water of greater
than $40\%$. The terrestrial planets in our solar system are thought
to have formed from water-devoid planetesimals. Hence the
planetesimals must have formed either before the snow line moved
inward or after the snow line moved outward past the Earth's
orbit. \cite{oka11} included ice opacity as well as the silicate
opacity and find similar results to \cite{garaud07}. They find that
there is a deficit of solid mass at the later times and formation of
water-devoid planetesimals is impossible. It is improbable for the
planetesimals to form before the snow line moves inwards because the
timescale involved is very short. \cite{machida10} investigated the
possibility that the Earth formed from sublimating icy planetesimals
after the snow line moved outwards. However, the extremely low
observed water content on Earth appears to be an unlikely outcome,
because of the competition between sublimation and collision of the
planetesimals. In this paper we attempt to address the inconsistency
between the theoretical models for the snow line evolution and the
observations of the solar system.

Protoplanetary discs may have a region of low turbulence at the disc
mid-plane. The MRI drives turbulence but may be suppressed by a low
ionisation fraction \citep{gammie96,gammie98}. The inner parts of the
disc are hot enough to be thermally ionised but farther away from the
central star, cosmic rays are the dominant source of ionisation and
these can only penetrate the surface layers. The mid-plane layer, the
dead zone, has no turbulence.  A variation of the turbulence with
height from the mid-plane has been considered in some previous snow
line models \citep[e.g.][]{kretke07}. However, the mid-plane layer had
a sufficiently large viscosity that a steady disc was still
achieved. \cite{kretke10} consider the vertical structure of a layered
disc and find that the location of the snow line is a little more
ambiguous with the vertical stratification. The upper layers of the
disc have a cooler temperature and so the snow line is closer to the
star in the surfaces than at the mid-plane.  For an accretion rate of
$10^{-8}\,\rm M_\odot\,yr^{-1}$ they predict a snow line radius of
around $1\,\rm AU$. This is similar to that found by \cite{garaud07}
and \cite{oka11} for the fully turbulent disc model.

If there is no turbulence within the dead zone, the disc cannot be in
a steady state and it is this possibility that we consider
here. Material accumulates in the dead zone which can then become
gravitationally unstable. While the infall accretion rate is high,
this can lead to the gravo-magneto disc instability. The disc spends
the majority of the time with a dead zone in a quiescent state with a
low accretion rate on to the star. However, the extra turbulence
driven by the gravitational instability in the dead zone increases the
disc temperature until the MRI is triggered. This causes a large
accretion outburst on to the star that is thought to explain FU
Orionis outbursts \citep{armitage01, zhu10, martin11}. At later times,
when the infall accretion rate drops, there is not sufficient material
in the disc for the gravo-magneto instability to operate but a dead
zone may still be present \citep{martin12a}. Given this rather
different (from a steady-state disc) configuration and behaviour, in
this paper, we consider the evolution of the snow line in a time
dependent disc model that includes a dead zone.

\section{Protoplanetary Disc Model}

The material in an accretion disc orbits the central mass, $M$, with a
Keplerian velocity at radius $R$ with angular velocity
$\Omega=\sqrt{GM/R^3}$ \citep{lyndenbell74,pringle81}. We use a one
dimensional layered disc model described in \cite{martin11} and
further developed in \cite{martin12a} to evolve the total surface
density, $\Sigma(R,t)$ and mid-plane temperature, $T_{\rm c}(R,t)$. We
take a solar mass star, $M=1\,\rm M_\odot$, with a disc that extends
from a radius of $R=5\,\rm R_\odot$ up to $R=40\,\rm AU$. Turbulence
is driven by the MRI in some regions of the disc and we parametrise
this viscosity with a \cite{shakura73} $\alpha$ parameter that we take
to be $0.01$ \citep[e.g.][]{brandenburg95,stone96} although there is
still some uncertainty in this value \citep[e.g.][]{king07}. The disc
consists of a turbulent surface layer of surface density, $\Sigma_{\rm
  m}$ and temperature $T_{\rm m}$.  Where it exists, the dead zone has
surface density $\Sigma_{\rm g}=\Sigma-\Sigma_{\rm m}$. The extent of
the dead zone in the disc is determined by a critical magnetic
Reynolds number, $Re_{\rm M,crit}(R,z)$. We use the analytical
approximations for the surface density in the turbulent layer given in
equations~26 and~27 of \cite{martin12b}.  The innermost parts of the
disc are externally heated by a flux of radiation from the central
star \citep[see also][]{chiang97}. We take the temperature of the star
$T_{\rm star}=3000\,\rm K$ and its radius to be $R_{\rm star}=2\,\rm
R_\odot$.

We consider a model for the collapse of a molecular cloud on to the
disc \citep{armitage01,martin12a}. Initially the accretion rate is
$2\times 10^{-5}\,\rm M_\odot\,yr^{-1}$ and this decays exponentially
on a timescale of $10^5 \,\rm yr$. Material is added to the disc at a
radius of $35\,\rm AU$. The dead zone can become self gravitating when
the Toomre parameter \citep{toomre64} is sufficiently small, $Q<2$.
We approximate the radius of the snow line to be where the temperature
of the disc is $T_{\rm snow}=170\,\rm K$. The disc models of
\cite{lecar06} suggest that this is a good approximation for varying
disc mass and opacity.

\begin{figure*}
\includegraphics[width=8.4cm]{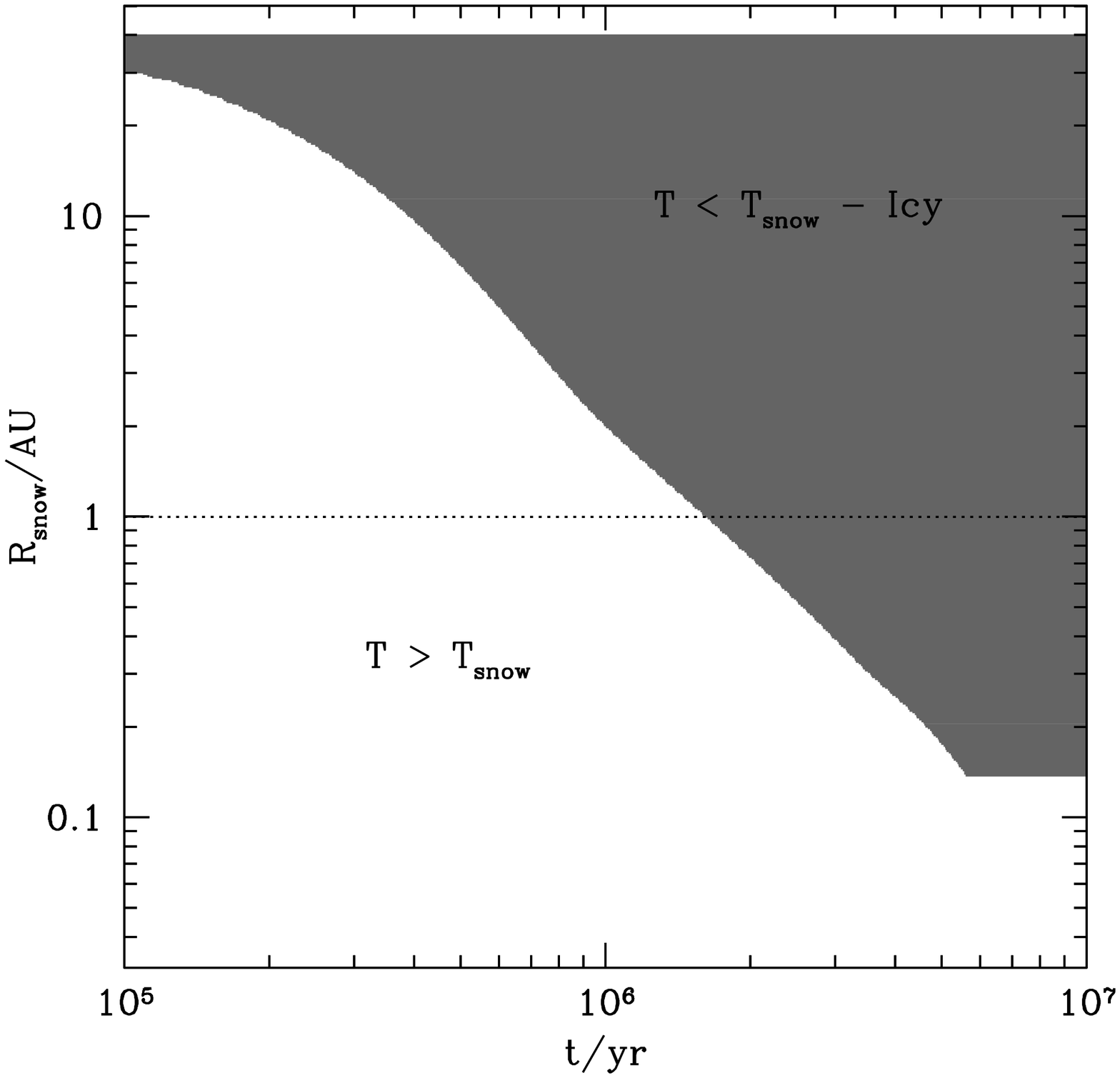}
\includegraphics[width=8.4cm]{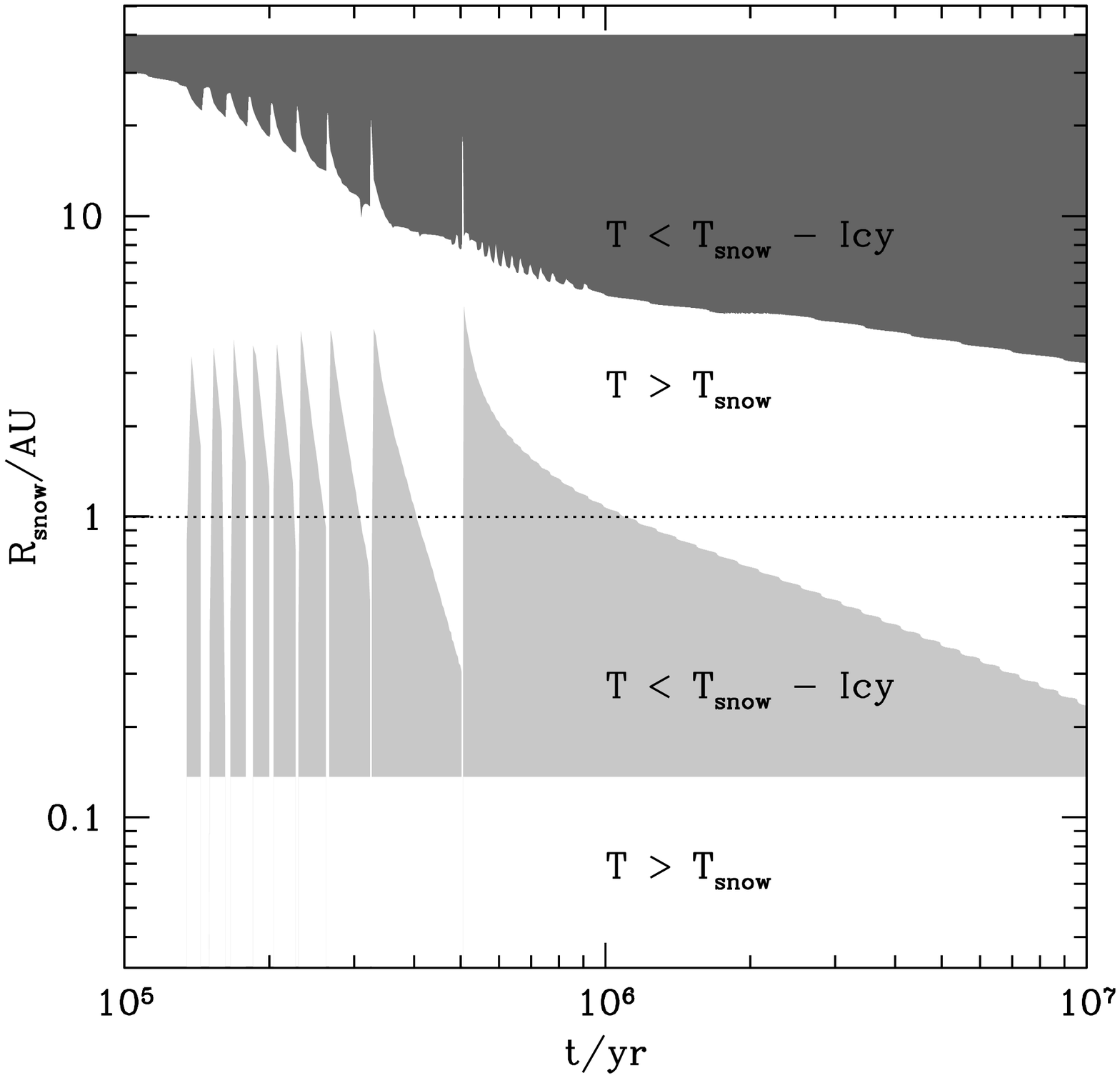}
\caption{The evolution of the icy regions of a protoplanetary disc
  around a solar mass star. Left: A fully turbulent disc ($Re_{\rm
    M,crit}=0$). Right: A disc with a dead zone determined by a
  critical magnetic Reynolds number of $Re_{\rm M,crit}=5\times 10^4$.
  Time begins at $t=10^5\,\rm yr$. The shaded areas show the icy
  regions of the disc. Where a dead zone exists there is an inner icy
  region (pale shaded areas) as well as the outer icy region (dark
  shaded areas). In the innermost parts of the disc, the radiation
  flux from the central star heats the disc above the snow line
  temperature, $T_{\rm snow}$. The dotted line shows the current
  radius of the Earth's orbit.}
\label{sl}
\end{figure*}

In Fig.~\ref{sl} we show the evolution of the snow line in two model
discs, one that is fully turbulent (or equivalently $Re_{\rm
  M,crit}=0$) and one that has a dead zone determined by a critical
magnetic Reynolds number of $Re_{\rm M,crit}=5\times 10^4$. The fully
turbulent model shows a monotonic decrease of the snow line radius in
time so that at late times the snow line is within the radius of the
Earth's orbit at $1\,\rm AU$. We do not obtain the increase of the
snow line radius at smaller accretion rates shown by \cite{oka11}
because there is sufficient mass in the disc that the model remains
optically thick at the radius where the snow line occurs. For example,
at a time of $2 \times 10^6\,\rm yr$, the infall accretion rate on to
the disc has dropped to less than $10^{-13}\,\rm M_\odot\,yr^{-1}$ but
the accretion on to the star is $6\times 10^{-9}\,\rm
M_\odot\,yr^{-1}$. The disc does not remain in a quasi-steady state as
the accretion rate decreases as assumed in previous models.

\begin{figure}
\includegraphics[width=8.9cm]{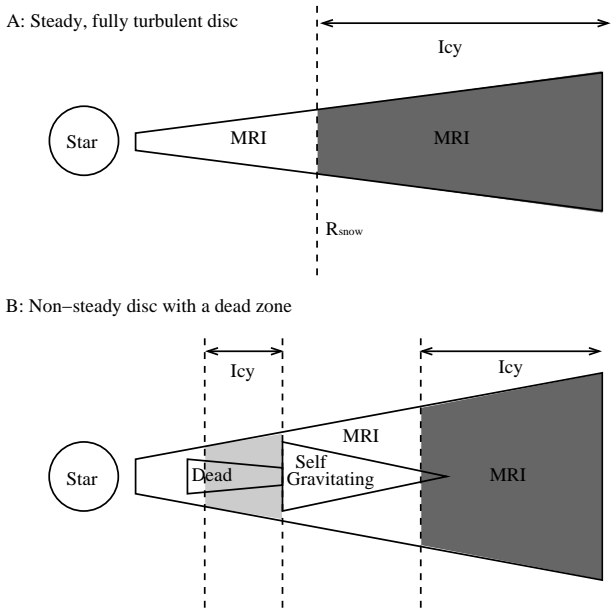}
\caption{Sketches of the disc structure for the two models. Figure~A
  represents a steady disc that is fully turbulent. The temperature
  decreases monotonically with radius and only the outer regions of
  the disc are icy. Figure~B represents a disc with a dead zone. The
  outer parts of the dead zone are self-gravitating, more massive and
  thus have a higher temperature.. There are two icy regions in this
  model. The outer icy region is farther out than that predicted by
  the fully turbulent model. The inner icy region exists within the
  dead zone.The shaded regions correspond to the inner (pale shaded)
  and outer (dark shaded) icy regions as described in
  Fig.~\ref{sl}.}
\label{fig}
\end{figure}

For the model with $Re_{\rm M,crit}=5\times 10^4\,\rm yr$, initially
there is no dead zone and the disc evolution proceeds as for the case
with $Re_{\rm M,crit}=0$. As the infall accretion rate drops, the disc
cools and a dead zone forms in the protoplanetary disc. The inner
parts of the disc remain hot because they are heated by the radiation
from the central star. Within the dead zone, an icy region exists. The
outer parts of the dead zone become self gravitating and this drives a
small amount of turbulence that heats the disc above the snow line
temperature (see the sketch of the two disc models in
Fig.~\ref{fig}). The very brief periods of time during which there is
no inner icy region are when the disc is in outburst. Self gravity
heats the disc sufficiently to trigger the MRI within the dead
zone. The temperature of the disc increases significantly and there is
an FU Orionis type outburst where the accretion rate on to the star
increases by several orders of magnitude \citep[see][]{martin12a}.

As with the fully turbulent disc, the outermost parts of the disc are
cool and icy. At late times the outer icy region reaches a minimum
radius of about $3.1\,\rm AU$, roughly consistent with current
observations.  In our solar system, the Earth and other terrestrial
planets likely formed after the inner icy region passed inside their
radius because they are water devoid. This means that the Earth must
have formed after a time of about $10^6\,\rm yr$. At this time, there
is little infall accretion on to the disc but there is a significant
amount of mass in the disc, a total of about $0.3\,\rm M_\odot$.  The
surface density at $R=1\,\rm AU$ is high at $7.5\times 10^4\,\rm
g\,cm^{-2}$. This is in agreement with the standard minimum mass solar
nebular at this radius of about $1.7\times 10^3\,\rm g\,cm^{-2}$
\citep{hayashi81}. It is comparable to the model of \cite{desch07}
based on the 'Nice' model of planet formation that predicts a surface
density of $5\times 10^4\,\rm g\,cm^{-2}$ at the radius of the Earth's
orbit. At this time there was still sufficient mass in the disc for
planetesimal and Earth mass planet formation from the water-devoid
disc.

\section{Discussion}

Our simplified model clearly contains a number of uncertainties. Note
that because we took the upper value for $T_{\rm snow}$, our results
show the upper limit of the extent of the icy regions. Also, we have
assumed, as did previous researchers, that the radius of the snow line
in the gas disc is equivalent to the radius of the snow line of the
planetesimal disc \citep[e.g.][]{oka11}. However, once the planetesimals
grow large enough, they may decouple from the gas disc, and follow
quasi-circular orbits. If they were to move over the snow line, then
they would sublimate or condense at the water-ice evaporation front
\citep{stevenson88}. The process of sublimation can occur on a short
timescale \citep[e.g.][]{sack93,brown07} but condensation is less well
understood. \cite{marseille11} find there is an extended region beyond
the snow line where both icy and bare grains coexist, for a radial
distance of about $0.4\,\rm au$. This should be investigated further
in future work with a disc model that tracks the evolution of the
water. While \cite{ciesla06} investigated this phenomenon in a fully
turbulent disc, such calculations should be extended to a disc that
contains a dead zone.

There are other unknown parameters in the model such as the
value of the critical magnetic Reynolds number
\citep[e.g.][]{fleming00}, and the viscosity parameter
$\alpha$. \cite{martin12a} showed that in order to reproduce FU
Orionis outbursts, $Re_{\rm M,crit}$ must be a few $10^4$. There is
also some uncertainty about whether the surface density in the
turbulent surface layer is determined by Ohmic resistivity (as we have
assumed) or if ambipolar diffusion or the Hall effect could play an
important role \citep[e.g.][]{perezbecker11,bai11}. On this last
point, however, \cite{wardle11} find the Ohmic resistivity term
provides an average value for the active layer surface density for a
range of vertical magnetic fields. We should also note that there
appears to be some inconsistency between the accretion rates predicted
by these dead zone models and observed T Tauri rates.  The work
presented here is not intended to be a comprehensive study of the
evolution of the snow line. Rather, the purpose is to show that the
formation a dead zone in a protoplanetary disc {\it can} keep the
outer snow line farther away from the central star. Thus, a model that
includes a dead zone can provide a solution to the problem that the
Earth must form from water-devoid planetesimals that may not be
present at $1\,\rm AU$ in a fully turbulent disc model. This should be
investigated further in future work with more detailed numerical
simulations once some of the current issues have been resolved.

On a more speculative note, the inner icy region, not predicted by the
turbulent disc model, could allow for the formation of icy planets or
gas giants in the inner regions of exo-solar systems. For example, at
a time of $6\times 10^5\,\rm yr$ there is a mass of $6.7\,\rm M_{\rm
  J}$ in the inner icy region of the disc. The cores of giant planets
are usually assumed to form in the cool region beyond the outer snow
line. For example, hot Jupiters are giant planets that are very close
to their central star. They are thought to have migrated through the
disc to their current position \citep[e.g.][]{masset03}. However, with
the proposed disc model (including a dead zone) it is possible that
they could form much closer to the star, and closer to their observed
locations. The newly formed core would not migrate inwards very fast
because it would be within a dead zone \citep[e.g.][]{matsumura07} and
the dead zone extends closer to the star than the icy region (see
Fig.~\ref{fig}). It is possible that the inner icy region could exist
for around $10^7\,\rm yr$ leaving perhaps sufficient time for giant
planet core formation.  However, we note that in order to form icy
planetesimals we need not only a low temperature, but also water
vapour to be present. For example, if the cold gas is dry, then icy
planetesimals will neither form, nor will ice be added to
pre-existing planetesimals. The speculative possibility of the formation of hot
Jupiters at their current location needs further investigation.

\section{Conclusions}

We have computed the evolution of the icy regions of a protoplanetary
disc that contains a dead zone. The disc evolves from times of high
infall accretion through FU Orionis outbursts to late times when the
infall accretion rate is negligible and planetesimals are thought to
form. In a fully turbulent disc, the unique snow line passes inside
the radius of the Earth's orbit causing problems for the formation of
our water-devoid planet Earth.  However, when a dead zone forms in the
disc, it prevents mass flow through the disc and the outer regions
become self-gravitating. The turbulence driven by self-gravity
increases the temperature of the outer parts of the dead zone and thus
the outer icy region is much farther from the star as shown in the
sketches in Fig.~\ref{fig}.  Our model also predicts an inner icy
region during early times of disc evolution, while a dead zone is
present. This moves inside the radius of the Earth's orbit after a
time of around $10^6\,\rm yr$.  The outer icy region exists at all
times but has a minimum radius of around $3.1\,\rm AU$, allowing for
the formation of our water-devoid planet at its current radius.

The key point of the present work is that the inclusion of a dead zone
in time-dependent protoplanetary disc models can significantly change
the evolution of the snow line. We have shown that it could resolve
the apparent contradiction of the formation of the Earth from
water-devoid planetesimals and it also introduces the possibility for
icy planet formation close to the central star.

\section*{Acknowledgements}

We thank Jim Pringle and an anonymous referee for useful comments. RGM
thanks the Space Telescope Science Institute for a Giacconi
Fellowship.

\label{lastpage}
\end{document}